\documentclass{revtex4}
\usepackage{eurosym}
\usepackage{amssymb}
\usepackage{amsmath}

\input{tcilatex}

\begin{document}

\title{Analytical solutions in rotating linear dilaton black holes: resonant
frequencies, quantization, greybody factor, and Hawking radiation}
\author{I. Sakalli}
\email{izzet.sakalli@emu.edu.tr}
\affiliation{Physics Department , Eastern Mediterranean University, Famagusta, Northern
Cyprus, Mersin 10, Turkey}
\date{\today }

\begin{abstract}
Charged massive scalar fields are studied in the gravitational,
electromagnetic, dilaton, and axion fields of rotating linear dilaton black
holes. In this geometry, we separate the covariant Klein--Gordon equation
into radial and angular parts and obtain the exact solutions of both the
equations in terms of the confluent Heun functions. Using the radial
solution, we study the problems of resonant frequencies, entropy/area
quantization, and greybody factor. We also analyze the behavior of the wave
solutions near the event horizon of the rotating linear dilaton black hole
and derive its Hawking temperature via the Damour--Ruffini--Sannan method.
\end{abstract}

\maketitle

\section{Introduction}

Analytical solutions to a wave equation in a black hole background (in
particular, for a stationary black hole) are of remarkable importance in
theoretical and mathematical physics. Obtaining an exact solution to the
wave equation is widely applicable in black hole physics. For example, one
can study the quasinormal modes \cite{Leav,Ono,Kokko,Hod}, analyze the
entropy and area quantization \cite{PRLBQ}, compute the greybody factor and
absorption rate of a black hole \cite{Staro1,Staro2,P1,P2,GUnruh,Suzuki,Harm}%
, and study black hole perturbation \cite{Chandra}. The behavior of scalar
fields in the backgrounds of black holes is studied to understand the
physics of spin-0 particles. Thus, it is important to seek analytical
solutions to the Klein--Gordon equation (KGE) and analyze physical phenomena
such as the emission of scalar particles from black holes. Therefore, exact
solutions of the KGE in various black hole geometries have been studied,
e.g., \cite{kg1,kg2,kg3,kg4,kg6,kg7,kg8,kg9,kg10,kg11,kg12,kg13,kg14,kg15}
and references therein.

The subject of quantization of black holes was introduced in the early 1970s
by Bekenstein \cite{Bek001,Bek002}. He conjectured that a black hole has an
equidistant area ($A_{BH}$) spectrum: $A_{BH\_n}=8\pi n\hbar ,$ $\
(n=0,1,2..)$ \cite{Bek003,Bek004,Bek005}. To this end, he used the
Ehrenfest's principle by considering the black hole area as an adiabatic
invariant quantity. After those seminal works of Bekenstein, numerous
theoretical models have been proposed for quantizing the black holes; for a
topical review see \cite{RevG}. Among them, the method of Maggiore \cite%
{Magg0} have gained much attention in the literature. According to Maggiore,
a black hole can be considered as a damped harmonic oscillator with a
frequency ($\omega $), which is identical to the complex resonant
frequencies or quasinormal modes. For the high damping oscillations, the
imaginary part of the resonant frequencies always becomes dominant over the
real part. Based on this fact Maggiore used the transition frequency $\Delta
\omega \approx $\textit{Im}$\omega _{n-1}-$\textit{Im}$\omega _{n}$ in the
adiabatic invariance formula of Kunstatter \cite{Kuns0} and proved that the
area spectrum is exactly equal to Bekenstein's original result \cite{Bek005}.

Hawking \cite{Hawrad1} proved that black holes can thermally create and emit
quantum particles, known as Hawking radiation, until they exhaust their
energy and evaporate completely. According to this theory, black holes are
therefore neither completely black nor do they last forever. In fact, the
Hawking radiation is a phenomenon in which both the general theory of
relativity and quantum theory simultaneously play an active role.
Especially, it can be seen as the onset of the theory of quantum gravity. In
the last four decades, various methods have been proposed to compute the
Hawking radiation, e.g., \cite%
{Hex1,Hex2,Hex3,Hex4,Hex40,Hex41,Hex5,Hex6,Hex61,Hex7,Hex8,Hex81}. The
numerous publications on this subject to date clearly show that the Hawking
radiation remains central in theoretical physics. Moreover, in experimental
physics, Steinhauer \cite{Stein} has almost succeeded in creating a
laboratory-scale imitation of a black hole that emits Hawking radiation;
that is, the particles that escape black holes because of quantum mechanical
effects. One of the most valuable contributions to the original derivation
of Hawking was made by Damour, Ruffini \cite{Hex2}, and Sannan \cite{Sannan}
(DRS). They demonstrated that for obtaining the decay or emission rate, it
is necessary to build a damped part in the outgoing wave function.
Therefore, one should apply a simple analytic continuation to the outgoing
wave function (available in the exterior region) for obtaining its internal
region structure. Such an analytical extension produces a damping factor and
yields the scattering probability (i.e., emission rate) of the scalar wave
at the event horizon. Furthermore, the DRS method only requires the
existence of a future horizon and is independent from the dynamical details
of the horizon formation process. For applications of the DRS method, the
reader may refer to \cite{kg14,kg15,DRS1,DRS11,DRS2,DRS3,DRS4,DRS5,DRS6}.
This method is also closely related to the near-horizon conformal property
of the black hole geometry \cite{CPbh,NAF1}. Although many studies of the
DRS method have focused on the asymptotically flat black holes, the
applications of the DRS method on the non-asymptotically flat black holes
have remained very limited \cite{NAF1,NAF2,NAF3}. Our present study aims to
decrease this paucity in the literature.

In this study, we mainly focus on the analytical solutions of the KGE for a
charged massive scalar field in rotating linear dilaton black holes (RLDBHs) 
\cite{Clem1}. These black holes represent non-asymptotically flat (like the
Friedmann--Lema\^{\i}tre--Robertson--Walker spacetime \cite{FLRW}, which is
the universe model that most theorists currently use)\ solutions to the
Einstein-Maxwel-Dilaton-Axion (EMDA) gravity theory. The experiments \cite%
{DME1,DME2,DME3,DME4} on dilaton and axion fields that naturally exist in
the RLDBH geometry may vindicate the dark matter in the near future. At the
present time, the studies for the RLDBH \cite%
{rbh0,rbh1,rbh2,rbh3,rbh4,rbh5,rbh52} in the literature are relatively less
than the studies that exist for static linear dilaton black holes \cite%
{rbh52,ldbh1,ldbh2,ldbh3,ldbh3n,ldbh4,ldbh5,ldbh6}. In particular, for the
problems of absorption cross-section and decay rate for the massless and
chargeless bosons emitted by a RLDBH, the reader is referred to \cite{rbh5}.
We show that the obtained charged massive scalar wave function solutions are
expressed in terms of the confluent Heun functions \cite%
{heun0,heun1,heun2,heun3}. The solutions cover the region between the outer
horizon and spatial infinity. Inspiring from the very recent study of Vieira
and Bezerra \cite{DRS6}, we derive the resonant frequencies of the RLDBH by
using the analytical solution of the radial equation and in sequel derive
the equally spaced entropy/area spectra of this black hole. We also study
the greybody problem of the RLDBH spacetime. But, the limited linear
transformation formulas of the confluent Heun functions lead us to consider
the case of chargeless and massless scalar fields. By obtaining the greybody
factor, we reveal which waves are eligible to travel from the horizon to the
asymptotic region in the RLDBH geometry. We then investigate the Hawking
emission of the chargeless and massless spin-0 particles by computing the
emission rate within the framework of the DRS method.

The paper is divided into the following sections. In Sec. II, we introduce
the metric of the RLDBH spacetime and demonstrate its thermodynamic
features. Section III is devoted to the KGE for charged massive scalar
fields in the RLDBH geometry. Moreover, we separate the KGE into angular and
radial parts. In Sec. IV, the analytical solutions of the angular and radial
equations are represented in terms of the confluent Heun functions. In Sec.
V, we present the applications of the wave solution. In this regard, the
problems of resonant frequencies, entropy/area quantization, greybody
factor, and Hawking radiation are elaborately studied. Finally, we summarize
our discussions in the conclusion section. (We use geometrized units where $%
G=c=1$, so that energy and time have units of length. Appendix lists the
prominent symbols that are used throughout the paper.)

\section{Rotating Linear Dilaton Black Hole Spacetime}

The EMDA theory is described by the following action \cite{Leygn} 
\begin{equation}
S=\frac{1}{16\pi }\int d^{4}x\sqrt{\left\vert g\right\vert }\bigg(\Re
-e^{-2\phi }F_{\mu \nu }F^{\mu \nu }-\varkappa F_{\mu \nu }\widetilde{F}%
^{\mu \nu }-2\partial _{\mu }\phi \partial ^{\mu }\phi -\frac{1}{2}e^{4\phi
}\partial _{\mu }\chi \partial ^{\mu }\chi \bigg),  \label{1n}
\end{equation}

where $\phi $ and $\chi $ represent the dilaton and axion fields,
respectively. The Maxwell field is governed by $F_{\mu \nu }=\partial _{\mu }%
\mathcal{A}_{\mathcal{\nu }}-\partial _{\mathcal{\nu }}\mathcal{A}_{\mu }$
in which $\mathcal{A}$ is an Abelian vector field (i.e., electromagnetic
vector potential), $\widetilde{F}^{\mu \nu }$ is the dual of $F_{\mu \nu }$,
and $\Re $ denotes the Ricci scalar. The line-element which is the solution
to Eq. (1) corresponds to the rotating linear dilaton metric with mass term, 
$M$, rotation parameter, $a$, and the background electric charge, $Q$. In
Boyer-Lindquist coordinates, the RLDBH spacetime \cite{Clem1} is given by

\begin{equation}
ds^{2}=-fdt^{2}+\frac{1}{f}dr^{2}+\xi \left[ d\theta ^{2}+\sin ^{2}\theta
\left( d\varphi -\frac{a}{\xi }dt\right) ^{2}\right] ,  \label{2n}
\end{equation}

where

\begin{equation}
f=\Delta \xi ^{-1},  \label{3}
\end{equation}%
\begin{equation}
\xi =rr_{0},  \label{4}
\end{equation}

and the horizon surface equation ($f=0$) is obtained from the condition

\begin{equation}
\Delta =(r-r_{+})(r-r_{-}),  \label{5}
\end{equation}

in which $r_{+}$ and $r_{-}$ represent the event and inner (Cauchy)
horizons, respectively. Those radii are given by

\begin{equation}
r_{\pm }=M\pm \sqrt{M^{2}-a^{2}},  \label{6}
\end{equation}

where $M$ is the twice of the quasilocal mass $\left( M=2M_{QL}\right) $ 
\cite{BY}. The charge parameter $r_{0}=\sqrt{2}Q$ and the rotation parameter 
$a$ are related with the angular momentum ($J$) as $2J=ar_{0}$ \cite{Clem1}.
Once the rotation ceases, the stationary metric (2) clearly becomes static 
\cite{Hex7}. Furthermore, the dilaton and axion fields are given by

\begin{equation}
e^{-2\phi }=\frac{\xi }{r^{2}+a^{2}\cos ^{2}\theta },  \label{7}
\end{equation}

\begin{equation}
\chi =-\frac{r_{0}a\cos \theta }{r^{2}+a^{2}\cos ^{2}\theta }.  \label{8}
\end{equation}

The electromagnetic vector potential is 
\begin{equation}
\mathcal{A=A}_{t}dt+\mathcal{A}_{\varphi }d\varphi =\frac{1}{\sqrt{2}}\left(
e^{2\phi }dt+a\sin ^{2}\theta d\varphi \right) ,  \label{9}
\end{equation}

and the Maxwell $2$-form is derived as follows:

\begin{equation}
F=\frac{1}{\sqrt{2}}\Bigg[\frac{r^{2}-a^{2}\cos ^{2}\theta }{r\xi }dr\wedge
dt+a\sin 2\theta \,d\theta \wedge \bigg(d\varphi -\frac{a}{\xi }dt\bigg)%
\Bigg].  \label{10}
\end{equation}

\subsection{Thermodynamics}

In this section, we discuss the thermodynamic features of the RLDBHs. First,
we consider a particle near the horizon with the following 4-velocity

\begin{equation}
\boldsymbol{u}=u^{t}\left( \partial _{t}+\frac{a}{\xi }\partial _{\varphi
}\right) ,  \label{11}
\end{equation}

which satisfies the normalization condition

\begin{equation}
1=u^{\mu }u_{\mu }.  \label{12}
\end{equation}

Hence, one finds (near the horizon)

\begin{equation}
u^{t}=\frac{1}{\sqrt{g_{tt}}}.  \label{13}
\end{equation}

Because all the metric components are only functions of $r$ and $\theta $,
particle acceleration can be obtained from

\begin{equation}
\mathit{a}^{\mathit{\mu }}=\Gamma _{\alpha \beta }^{\mu }u^{\alpha }u^{\beta
}=-g^{\mu \alpha }\partial _{\alpha }\ln u^{t}.  \label{14}
\end{equation}

The definition of the surface gravity ($\kappa $) is given by \cite{WaldGR}

\begin{equation}
\kappa =\lim_{r\rightarrow r_{+}}\frac{\sqrt{\mathit{a}^{\mathit{\mu }}%
\mathit{a}_{\mathit{\mu }}}}{u^{t}}.  \label{15}
\end{equation}

After a straightforward calculation, Eq. (15) becomes

\begin{equation}
\kappa =\frac{1}{2}\left. \frac{df}{dr}\right\vert _{r=r_{+}}=\frac{%
r_{+}-r_{-}}{2r_{+}r_{0}}.  \label{16}
\end{equation}

Thus, we obtain the Hawking temperature as

\begin{equation}
T_{H}=\frac{\hslash \kappa }{2\pi }=\frac{\hslash \left( r_{+}-r_{-}\right) 
}{4\pi r_{+}r_{0}}.  \label{17}
\end{equation}

The angular velocity ($\Omega _{H}$) and the black hole area are given by

\begin{equation}
\Omega _{H}=-\left. \frac{g_{_{t\varphi }}}{g_{\varphi \varphi }}\right\vert
_{r=r_{+}}=\frac{a}{r_{+}r_{0}},  \label{18}
\end{equation}

\begin{equation}
A_{BH}=\int_{0}^{2\pi }d\varphi \int_{0}^{\pi }\sqrt{-g}\ d\theta =4\pi
r_{+}r_{0}.  \label{19}
\end{equation}%
It is worth noting that in the evaluation of integral (19), $\sqrt{-g}$ is
considered for the metric tensor of the RLDBH horizon (setting $dr=dt=0$):

\begin{equation}
g_{\mu \nu }=\left( 
\begin{array}{cc}
r_{+}r_{0} & 0 \\ 
0 & r_{+}r_{0}\sin ^{2}\theta%
\end{array}%
\right) .  \label{20}
\end{equation}

Hence, the entropy of the black hole, $S_{BH}$, is given by 
\begin{equation}
S_{BH}=\frac{A_{BH}}{4\hslash }=\frac{\pi r_{+}r_{0}}{\hslash }.  \label{21}
\end{equation}

The quantities described by Eqs. (17), (18), and (21) satisfy the first law
of thermodynamics: 
\begin{equation}
dM_{QL}=T_{H}dS_{BH}+\Omega _{H}\ dJ.  \label{22}
\end{equation}

\section{Separation of KGE in RLDBH geometry}

In this section, we consider the wave equation of the charged massive scalar
particles propagating in the geometry of RLDBH.

The KGE for a charged massive scalar particle is given by (e.g., \cite{PLB}) 
\begin{equation}
\frac{1}{\sqrt{-g}}\left( \partial _{\alpha }-iq\mathcal{A}_{\alpha }\right)
\left( \sqrt{-g}g^{\alpha \nu }(\partial _{\nu }-iq\mathcal{A}_{\nu })\Psi
\right) -\mu _{s}^{2}\Psi =0,  \label{23}
\end{equation}%
where $\mu _{s}$ and $q$ represent the mass and charge of the scalar field $%
\Psi $, respectively. Owing to the axial symmetry and time independence of
the spacetime, the scalar field can be written as 
\begin{equation}
\Psi =\Psi (\boldsymbol{r},t)=R(r)S(\theta )\mbox{e}^{im\varphi }\mbox{e}%
^{-i\omega t}\ ,  \label{24}
\end{equation}%
where $\omega $ is the energy (frequency) corresponding to the flux of
particles at spatial infinity and $m$ denotes the azimuthal quantum number.
Thus, Eq. (23) takes the following form in the RLDBH spacetime 
\begin{eqnarray}
&&\frac{1}{2\xi f}\left[ {q}^{2}\left( {r}^{2}+{a}^{2}\right) ^{2}-\sqrt{2}q{%
a}m({a}^{2}+{r}^{2})+2{m}^{2}{a}^{2}\right] +\frac{\omega }{f}\bigg[\omega
\xi +\sqrt{2}q\left( {r}^{2}+{a}^{2}\right) -2am\bigg]+\sqrt{2}qma  \notag \\
&&-\xi \mu _{s}^{2}+\frac{1}{R\left( r\right) }\frac{d}{dr}\left[ \xi f{%
\frac{d}{dr}}R\left( r\right) \right] -{\frac{{m}^{2}}{\sin ^{2}\theta }}-%
\frac{1}{2}{q}^{2}{a}^{2}\sin ^{2}\theta +\frac{1}{\sin \theta S\left(
\theta \right) }\frac{d}{d\theta }\left[ \cos \theta {\frac{d}{d\theta }}%
S\left( \theta \right) \right] =0,  \label{25nn}
\end{eqnarray}

and by using an eigenvalue $\lambda $ one can separate Eq. (25) into an
angular equation 
\begin{equation}
{\frac{d^{2}}{d{\theta }^{2}}}S\left( \theta \right) +\cot \theta \left( {%
\frac{d}{d\theta }}S\left( \theta \right) \right) -\left( \frac{{m}^{2}}{%
\sin ^{2}\theta }+\frac{1}{2}{q}^{2}{a}^{2}\sin ^{2}\theta -\lambda \right)
S\left( \theta \right) =0,  \label{26}
\end{equation}%
and a radial equation 
\begin{eqnarray}
&&\frac{d}{dr}\left[ \xi f\left( {\frac{d}{dr}}R\left( r\right) \right) %
\right] +\biggl\{\frac{1}{2\xi f}-\sqrt{2}q{a}m({a}^{2}+{r}^{2})+2{m}^{2}{a}%
^{2}+\frac{\omega }{f}\left[ \xi \omega +\sqrt{2}q\left( {r}^{2}+{a}%
^{2}\right) -2am\right]  \notag \\
&&+\sqrt{2}qma-\xi \mu _{s}^{2}\biggr\}R\left( r\right) =0.  \label{27}
\end{eqnarray}

As demonstrated in the following sections, the above separations enable us
to find the solutions of the angular and radial equations in terms of the
confluent Heun functions. In particular, the solution of Eq. (27) can help
us compute the standard Hawking radiation of the RLDBH.

\section{Analytical solutions of the angular and radial equations}

In this section, we discuss the exact solutions of the angular and radial
parts of the KGE.

\subsection{Angular equation}

By changing the independent variable $\theta $ to a new variable $y$ 
\begin{equation}
\theta =\cos ^{-1}\left( 1-2y\right) .  \label{28}
\end{equation}%
equation (26) transforms into 
\begin{equation}
{\frac{d^{2}}{d{y}^{2}}}S\left( y\right) +{\frac{2\,y-1}{y\left( y-1\right) }%
\frac{d}{dy}}S\left( y\right) -\frac{1\,}{4}\Bigg[\left( \frac{{m}}{y\left(
y-1\right) }\right) ^{2}+\frac{4\lambda }{y\left( y-1\right) }+8{q}^{2}{a}%
^{2}\Bigg]S\left( y\right) =0.  \label{29n}
\end{equation}%
We also introduce a new function $H(y)$ via%
\begin{equation}
S(y)=\mbox{e}^{\tau y}\left( \frac{y}{y-1}\right) ^{m}H(y)\ ,  \label{30}
\end{equation}%
where 
\begin{equation}
\tau =2aq^{\ast },  \label{31}
\end{equation}%
in which $q^{\ast }=\frac{q}{\sqrt{2}}$. Function $H(y)$ satisfies the
following equation 
\begin{equation}
{\frac{d^{2}}{d{y}^{2}}}H\left( y\right) +\left[ 2\tau +\frac{2}{y-1}+{\frac{%
m-1}{y\left( y-1\right) }}\right] {\frac{d}{dy}}H\left( y\right) +{\frac{%
\left( 2y-1+m\right) \tau -\lambda }{y\left( y-1\right) }}H\left( y\right)
=0,  \label{32}
\end{equation}

which can be rewritten as the confluent Heun equation \cite{heun3} 
\begin{equation}
{\frac{d^{2}}{d{y}^{2}}}H\left( y\right) +\left( \widetilde{\alpha }+\frac{%
\widetilde{\beta }+1}{y}+\frac{\widetilde{\gamma }+1}{y-1}\right) {\frac{d}{%
dy}}H\left( y\right) +\left( \frac{\widetilde{\mu }}{y}+\frac{\widetilde{\nu 
}}{y-1}\right) H\left( y\right) =0.  \label{33n}
\end{equation}

The parameters $\widetilde{\alpha },$ $\widetilde{\beta },$ and $\widetilde{%
\gamma }$ are given by

\begin{equation}
\widetilde{\alpha }=2\tau ,\ \ \widetilde{\beta }=-\widetilde{\gamma }=-m.
\label{34}
\end{equation}

By setting

\begin{equation}
\widetilde{\eta }=\frac{m^{2}}{2}-\lambda ,\ \ \widetilde{\delta }=0,
\label{35n}
\end{equation}

the other two parameters $\widetilde{\mu }$ and $\widetilde{\nu }$ in Eq.
(32) become

\begin{equation}
\widetilde{\mu }=\frac{1}{2}(\widetilde{\alpha }-\widetilde{\beta }-%
\widetilde{\gamma }+\widetilde{\alpha }\widetilde{\beta }-\widetilde{\beta }%
\widetilde{\gamma })-\widetilde{\eta }=\tau \left( 1-m\right) +\lambda ,
\label{36n}
\end{equation}

\begin{equation}
\widetilde{\nu }=\frac{1}{2}(\widetilde{\alpha }+\widetilde{\beta }+%
\widetilde{\gamma }+\widetilde{\alpha }\widetilde{\gamma }+\widetilde{\beta }%
\widetilde{\gamma })+\widetilde{\delta }+\widetilde{\eta }=\tau \left(
1+m\right) -\lambda .  \label{37n}
\end{equation}

The solution of Eq. (32) is given by \cite{map18}

\begin{equation}
H\left( y\right) =C_{1}\mbox{HeunC}(\widetilde{\alpha },\widetilde{\beta },%
\widetilde{\gamma },\widetilde{\delta },\widetilde{\eta };y)+C_{2}y^{-\beta }%
\mbox{HeunC}(\widetilde{\alpha },-\widetilde{\beta },\widetilde{\gamma },%
\widetilde{\delta },\widetilde{\eta };y),  \label{38n}
\end{equation}

where $C_{1}$ and $C_{2}$ are the integral constants. Thus, the general
exact solution of the angular part (28) of the KGE for a charged massive
scalar field in the RLDBH geometry and over the entire range $0\leq y<\infty 
$ is 
\begin{equation}
S(y)=\mbox{e}^{\tau y}\left( \frac{y}{y-1}\right) ^{m}\Bigg[C_{1}\mbox{HeunC}%
\left( \widetilde{\alpha },\widetilde{\beta },\widetilde{\gamma },\widetilde{%
\delta },\widetilde{\eta };y\right) +C_{2}y^{-\beta }\mbox{HeunC}\left( 
\widetilde{\alpha },-\widetilde{\beta },\widetilde{\gamma },\widetilde{%
\delta },\widetilde{\eta };y\right) \Bigg].  \label{39}
\end{equation}

\subsection{Radial equation}

We follow the procedure described in Sec. (IV A) to show that the radial
equation (27) can also be transformed into the confluent Heun equation (33).
Thus, we first set 
\begin{equation}
z=\frac{r-r_{+}}{r_{-}-r_{+}},  \label{40}
\end{equation}%
and using this new coordinate, Eq. (27) transforms to 
\begin{eqnarray}
&&\frac{d^{2}}{d{z}^{2}}R\left( z\right) +{\frac{\left( 1-2z\right) {\frac{d%
}{dz}}R\left( z\right) }{z\left( 1-z\right) }}+\Bigg\{{\frac{q^{\ast
2}\left( {r_{+}}-{r_{-}}\right) ^{2}{z}^{2}}{\left( 1-z\right) ^{2}}}-{\frac{%
\left( 2q^{\ast }\omega \,{r_{0}}+4\,q^{\ast 2}{r_{+}}-{r_{0}}\mu
_{s}^{2}\right) \left( {r_{+}}-{r_{-}}\right) z}{\left( 1-z\right) ^{2}}} 
\notag \\
&&+{\frac{6q^{\ast }{r_{0}}\,{r_{+}}\omega \,+{r}_{{0}}^{{2}}{\omega }%
^{2}+\mu _{s}^{2}\left( {r_{-}}-2\,{r_{+}}\right) {r_{0}}+2\left( 3r_{+}^{2}+%
{a}^{2}\right) q^{\ast 2}-\lambda }{\left( 1-z\right) ^{2}}}-\frac{2}{\left( 
{r_{+}}-{r_{-}}\right) \left( 1-z\right) ^{2}z}\,  \notag \\
&&\times \Bigg[2q^{\ast }\bigg(3r_{+}^{2}\omega {r_{0}}-ma{r_{+}}+a(a\omega {%
r_{0}}-m{r_{-}})\bigg)+4q^{\ast 2}r_{+}^{3}-{r_{0}}\mu
_{s}^{2}r_{+}^{2}+\left( \mu _{s}^{2}{r_{0}r_{-}}+4{a}^{2}q^{\ast 2}-\lambda
+2{r_{0}}^{2}{\omega }^{2}\right) {r_{+}}  \notag \\
&&+\lambda {r_{-}}-2{r_{0}}\omega am\Bigg]+\frac{\left( r_{+}^{2}+{a}%
^{2}\right) ^{2}}{\left( {r_{+}}-{r_{-}}\right) ^{2}{z}^{2}\left( 1-z\right)
^{2}}\left( q^{\ast }-{\frac{\left( ma-\omega {r_{0}r_{+}}\right) }{%
r_{+}^{2}+{a}^{2}}}\right) ^{2}\Bigg\}R(z)=0.  \label{41n}
\end{eqnarray}

Moreover, when we apply a particular s-homotopic transformation \cite{kg15}
to the dependent variable $R(z)\rightarrow U(z)$, where 
\begin{equation}
R(z)=\mbox{e}^{\beta _{1}z}z^{\beta _{2}}(1-z)^{\beta _{3}}U(z)\ ,
\label{42}
\end{equation}%
and coefficients $\beta _{1}$, $\beta _{2}$, and $\beta _{3}$ are given by 
\begin{equation}
\beta _{1}=iq^{\ast }\left( r_{+}-r_{-}\right) ,  \label{43}
\end{equation}%
\begin{equation}
\beta _{2}={\frac{i\left[ \omega r_{+}r_{0}-ma+q^{\ast }\left( r_{+}^{2}+{a}%
^{2}\right) \right] }{r_{+}-r_{-}}},  \label{44}
\end{equation}%
\begin{equation}
\beta _{3}={\frac{i\left[ \omega r_{-}r_{0}-ma+q^{\ast }\left( r_{-}^{2}+{a}%
^{2}\right) \right] }{r_{+}-r_{-}}}.  \label{45}
\end{equation}%
Function $U(z)$ satisfies the confluent Heun equation (33) with the
following parameters:

\begin{equation}
\widetilde{\alpha }=2\beta _{1},\ \ \widetilde{\beta }=2\beta _{2},\ \ 
\widetilde{\gamma }=2\beta _{3}.  \label{46}
\end{equation}

\begin{equation}
\widetilde{\delta }=\left( r_{+}-r_{-}\right) \left[ r_{0}\mu
_{s}^{2}-2q^{\ast 2}\left( r_{+}+r_{-}\right) -2q^{\ast }\omega r_{0}\right]
,  \label{47}
\end{equation}

\begin{eqnarray}
&&\widetilde{\eta }=\frac{1}{{({r_{+}}-{r_{-}})^{2}}}\Bigg[2q^{\ast 2}\left(
r_{+}^{4}-2r_{+}^{3}{r_{-}}-{a}^{4}-2{a}^{2}{r_{+}r_{-}}\right) +2q^{\ast }%
\bigg (ma({r_{+}^{2}+r_{-}^{2}+2\,{a}^{2})}  \notag \\
&&{-{\omega {r_{0}}}({r_{-}}\,{a}^{2}}+{3\,{r_{-}}\,r_{+}^{2}-r_{+}^{3}+{a}%
^{2}{r_{+})}\bigg)}-{2(ma-\omega \,{r_{0}}\,{r_{+}})(ma-\omega \,{r_{0}}\,{%
r_{-}})}\Bigg]{{-\lambda -{r_{0}}\mu _{s}^{2}{r_{+}}}.}  \label{48n}
\end{eqnarray}

We recall that

\begin{equation}
\widetilde{\mu }=\frac{1}{2}(\widetilde{\alpha }-\widetilde{\beta }-%
\widetilde{\gamma }+\widetilde{\alpha }\widetilde{\beta }-\widetilde{\beta }%
\widetilde{\gamma })-\widetilde{\eta },  \label{49}
\end{equation}

\begin{equation}
\widetilde{\nu }=\frac{1}{2}(\widetilde{\alpha }+\widetilde{\beta }+%
\widetilde{\gamma }+\widetilde{\alpha }\widetilde{\gamma }+\widetilde{\beta }%
\widetilde{\gamma })+\widetilde{\delta }+\widetilde{\eta }.  \label{50}
\end{equation}

Thus, using solution (38) of the confluent Heun equation, the general
solution of Eq. (40) in the exterior region of the event horizon ($0\leq
z<\infty $) is given by 
\begin{equation}
R(z)=\mbox{e}^{\beta _{1}z}(1-z)^{\beta _{3}}\Bigg[C_{1}z^{\beta _{2}}%
\mbox{HeunC}(\widetilde{\alpha },\widetilde{\beta },\widetilde{\gamma },%
\widetilde{\delta },\widetilde{\eta };z)+C_{2}z^{-\beta _{2}}\mbox{HeunC}(%
\widetilde{\alpha },-\widetilde{\beta },\widetilde{\gamma },\widetilde{%
\delta },\widetilde{\eta };z)\Bigg],  \label{51n}
\end{equation}%
where $C_{1}$ and $C_{2}$ are constants.

Around the regular singular point $u=0$, the confluent Heun function \cite%
{heun1,heun2,heun3} behaves as 
\begin{eqnarray}
&&\mbox{HeunC}(\widetilde{\alpha },\widetilde{\beta },\widetilde{\gamma },%
\widetilde{\delta },\widetilde{\eta };u)=1+u\frac{\bigg(\widetilde{\beta }%
\left( \widetilde{\gamma }-\widetilde{\alpha }+1\right) +2\widetilde{\eta }-%
\widetilde{\alpha }+\widetilde{\gamma }\bigg)}{2(1+\widetilde{\beta })}+%
\frac{u^{2}}{8(1+\widetilde{\beta })(2+\widetilde{\beta })}  \notag \\
&&\times \Bigg(\widetilde{\beta }^{2}(\widetilde{\alpha }-\widetilde{\gamma }%
)^{2}-4\widetilde{\eta }\widetilde{\beta }(\widetilde{\alpha }-\widetilde{%
\gamma })+2\widetilde{\alpha }\widetilde{\beta }(2\widetilde{\alpha }-%
\widetilde{\beta }-3\widetilde{\gamma })+4\widetilde{\beta }\widetilde{%
\gamma }\left( \widetilde{\beta }+\widetilde{\gamma }\right) +3\left( 
\widetilde{\alpha }^{2}+\widetilde{\beta }^{2}\right)  \notag \\
&&+4\widetilde{\eta }\bigg(\widetilde{\eta }-2\left( \widetilde{\alpha }-%
\widetilde{\beta }-\widetilde{\gamma }-1\right) \bigg)-4\widetilde{\alpha }%
\left( \widetilde{\beta }+\widetilde{\gamma }\right) +4\widetilde{\beta }%
\widetilde{\delta }+\widetilde{\gamma }\left( 10\widetilde{\beta }+3%
\widetilde{\gamma }^{2}\right) +4\left( \widetilde{\beta }+\widetilde{\delta 
}+\widetilde{\gamma }\right) \Bigg)+...\ ,  \label{52nn}
\end{eqnarray}%
which is an auxiliary mathematical expression in the analysis of the Hawking
radiation.

\section{Applications of the wave solution}

In this section, we firstly study the resonant frequencies of the charged
and massive scalar waves propagating in the RLDBH geometry by using the
solution (51). Next, the obtained resonant frequencies are applied in the
rotational adiabatic invariant quantity \cite{rbh2,Vagenas,Medved} to derive
the quantum entropy/area spectra of the RLDBH. This problem is nothing but
the extension of \cite{rbh2} in which the massless and chargeless scalar
waves were considered. Then, we obtain the reflection coefficient and
greybody factor for particular scalar waves. Finally, we apply the DRS\
method to compute the Hawking radiation of the RLDBH.

\subsection{Resonant frequencies and spectroscopy of RLDBH}

In this subsection, we follow the recently developed technique of \cite{DRS6}
for computing the resonant frequencies for charged massive scalar waves
propagating in the RLDBH background. By using those frequencies, we show how
one can derive the equally spaced entropy/area spectra of the RLDBH.

Resonant frequencies are the proper modes at which a black hole freely
oscillates when excited by a perturbation. In fact, the resonant frequencies
are said quasinormal modes, in contrast to the normal modes of Newtonian
gravity, because they are damped by the emission of gravitational waves; as
a consequence, the corresponding eigenfrequencies are complex \cite{Ferrari}%
. The imaginary component of the frequency tells us how quickly the
oscillation will die away. So if one sets a black hole to vibrating, it
radiates the energy away in gravitational waves. Both rotating and static
black hole solutions have not just one resonant frequency but a whole series
of resonant frequencies (see for example \cite{FNbook} and references
therein). The resonant frequencies are associated with the radial solution
(51) for the certain boundary conditions. The solution should be finite on
the horizon and well behaved at asymptotic infinity. The latter remark
requires that $R(z)$ must have a polynomial form, which is possible with the 
$\widetilde{\delta }_{n}$-condition \cite{DRS6,heun3}: 
\begin{equation}
1+\frac{\widetilde{\beta }+\widetilde{\gamma }}{2}+\frac{\widetilde{\delta }%
}{\widetilde{\alpha }}=-n\ ,\text{ \ \ \ \ \ \ \ \ \ \ \ \ }n=0,1,2,\ldots \
.  \label{53n}
\end{equation}

By using Frobenius method and putting the power series expansion into the
confluent Heun's differential equation (33), the three-term recursive
relation of coefficients starts to appear. For obtaining the Heun
polynomials, one should impose the condition of $\mathbf{C}_{n+2}=0$, where $%
\mathbf{C}_{n}$ is one of the elements of the three-term recurrence relation 
\cite{heun3}. In fact, $\mathbf{C}_{n+2}=0$ is equivalent to the $\widetilde{%
\delta }_{n}$-condition (for the technical details, we refer the interested
reader to \cite{poly} and the references therein). In particular, Fiziev 
\cite{Fiziev1}\ showed that the confluent Heun's polynomials obtained from
the $\widetilde{\delta }_{n}$-condition (53) admit the most general class of
solutions to the Teukolsky master equation, which is correspondent with the
Teukolsky-Starobinsky identities (TSIs) \cite{TSI}. Please be reminded that
TSIs are the key element of the theory of perturbations to a gravitational
field (including the subject of quasinormal modes \cite{Fiziev2}) of
rotating relativistic objects. Detailed studies on the TSIs can be seen in
the seminal works of Chandrasekhar \cite{Chandra,ChandraS}.

From Eq. (53), we get 
\begin{equation}
1+i\frac{\left[ 4\left( r_{+}^{2}+{a}^{2}\right) q^{\ast 2}+4q^{\ast }\left(
\omega r_{0}r_{+}-ma\right) -\mu _{s}^{2}r_{0}\,\left( r_{+}-r_{-}\right) %
\right] }{2q^{\ast }\left( r_{+}-r_{-}\right) }=-n,  \label{54n}
\end{equation}%
Similar to the very recent work \cite{DRS6}, one can find an analytic
expression for the resonant frequencies $\omega _{n}$ from Eq. (54) as
follows 
\begin{equation}
\omega _{n}=m\Omega _{H}+q\Phi _{e}+2\kappa r_{0}\frac{\mu _{s}^{2}}{q^{\ast
}}+i(n+1)\kappa ,  \label{55}
\end{equation}

where%
\begin{eqnarray}
\Phi _{e} &=&-\frac{r_{+}^{2}+{a}^{2}}{\sqrt{2}r_{+}r_{0}},  \notag \\
&=&-\left. \mathcal{A}_{t}\right\vert _{r=r_{+},\theta =\left( 0,\pi \right)
},  \label{56}
\end{eqnarray}%
which is the electric potential of the RLDBH measured at the north/south
poles of the 2-sphere with radius $r_{+}$ \cite{Geoffrey}. From Eq. (55),
one can get the transition frequency between two highly damped ($%
n\rightarrow \infty $) neighboring states as follows

\begin{eqnarray}
\Delta \omega &\approx &\text{\textit{Im}}\omega _{n-1}-\text{\textit{Im}}%
\omega _{n},\text{ \ \ \ \ \ \ (}n\rightarrow \infty \text{)}  \notag \\
&=&\kappa =\frac{2\pi T_{H}}{\hbar }.  \label{57n}
\end{eqnarray}

For a black hole system with total energy $E$, the natural adiabatic
invariant quantity $I_{adb}$ is given by \cite{rbh2,Vagenas,Medved}

\begin{equation}
I_{adb}=\int \frac{dE}{\Delta \omega }\equiv \int \frac{T_{H}dS_{BH}}{\Delta
\omega }.  \label{58nn}
\end{equation}

For large quantum numbers ($n\rightarrow \infty $), the Bohr--Sommerfeld
quantization condition \cite{BSQ} applies and $I_{adb}$ acts as a quantized
quantity ($I_{adb}\simeq n\hbar $) \cite{Kuns0}. Inserting the transition
frequency (57) into Eq. (58), one finds

\begin{equation}
I_{adb}=\frac{\hbar S_{BH}}{2\pi }=n\hbar .  \label{59nn}
\end{equation}

From above, we read the entropy spectrum as

\begin{equation}
S_{BH\_n}=2\pi n.  \label{60nn}
\end{equation}

Since $S_{BH}=\frac{A_{BH}}{4\hbar },$ the area spectrum is then obtained as

\begin{equation}
A_{BH\_n}=8\pi n\hbar ,  \label{61nn}
\end{equation}

and the minimum change in the area becomes

\begin{equation}
\Delta A_{BH}^{\min }=8\pi \hbar .  \label{62nn}
\end{equation}

As it can be seen from Eqs. (60) and (61), both entropy and area spectra are
equally spaced and independent of the black hole parameters. Besides, Eq.
(62) shows that RLDBH\ horizon is made by patches of equal area $8\pi \hbar
. $ Namely, the results obtained are fully in agreement with the Bekenstein
conjecture \cite{Bek005} and with Ref. \cite{rbh2}. \ 

\subsection{Greybody factor and Hawking radiation}

The greybody factor accounts for the fact that waves need to travel from the
horizon to spatial infinity in the curved geometry \cite{Geoffrey}.
Analytical greybody factor computations require to know the behavior of the
general radial solution (51) near spatial infinity $r\rightarrow \infty $ or 
$z\rightarrow \infty $. To this end, there is a need for a transformation
similar to the following

\begin{equation}
\mbox{HeunC}\left( \widetilde{\alpha },\widetilde{\beta },\widetilde{\gamma }%
,\widetilde{\delta },\widetilde{\eta };z\right) \rightarrow \text{Gamma
functions}\times \mbox{HeunC}(\widetilde{a},\widetilde{b},\widetilde{c},%
\widetilde{d},\widetilde{e};\frac{1}{z}).  \label{63n}
\end{equation}

The parameters $\widetilde{a},\widetilde{b},\widetilde{c},\widetilde{d},%
\widetilde{e}$ and the Gamma functions introduced in the above equation
should be related with $\widetilde{\alpha },\widetilde{\beta },\widetilde{%
\gamma },\widetilde{\delta },\widetilde{\eta }$ according to the
transformation rules of the special functions \cite{SFbook}. The key point
here is the normalization condition \cite{heun1}: $\mbox{HeunC}(\widetilde{a}%
,\widetilde{b},\widetilde{c},\widetilde{d},\widetilde{e};\frac{1}{z}=0)=1$
while $z\rightarrow \infty $. Thus, the asymptotic solution of the
transformed [via Eq. (63)] radial equation (51) would describe the pure
asymptotic ingoing and outgoing waves [like the wave solution illustrated in
Eq. (75)], which allow us to evaluate original flux coming from infinity and
compare it to the flux at the black hole horizon. Then, the calculation of
the greybody factor would be possible for the RLDBH. But unfortunately such
a transformation (63) currently does not exist in the literature. In fact,
unlike many classical hypergeometric transformations \cite{MyBook}, the
confluent Heun functions have very limited transformations \cite{AHC}.
Nevertheless, we are not completely helpless. The following transformation 
\cite{map18} enables us to transform the confluent Heun functions to the
hypergeometric functions. 
\begin{equation}
\mbox{HeunC}(\widetilde{\alpha },\widetilde{\beta },\widetilde{\gamma },%
\widetilde{\delta },\widetilde{\eta };z)=\left( 1-z\right) ^{-\Xi }{F\Bigg(%
\Xi ,\Xi -}\widetilde{{\gamma }}{;\,1+\widetilde{{\beta }};\,{\frac{z}{z-1}}}%
\Bigg),\text{ \ \ \ \ \ }\left( \widetilde{\alpha }=0,\widetilde{\beta }%
+1\neq 0,\widetilde{\delta }=0,z\neq 1\right) ,  \label{64n}
\end{equation}

where

\begin{equation}
\Xi =\frac{1+\widetilde{\beta }+\widetilde{{\gamma }}{+}\sqrt{\widetilde{{%
\beta }}^{2}+\widetilde{{\gamma }}^{2}+1-4\widetilde{\eta }}}{2}.
\label{65n}
\end{equation}

However, the conditions of $\widetilde{\alpha }=0$ and $\widetilde{\delta }%
=0 $\ for non-extremal ($r_{+}\neq r_{-}$) RLDBH are simultaneously
satisfied if and only if the chargeless ($q=0$ $\rightarrow \beta _{1}=0$)
and massless ($\mu _{s}=0$) scalar fields are taken into account [see Eqs.
(43), (46), and (47)]. In this case, the general radial solution (51)
reduces to the following form

\begin{equation}
R(z)=C_{1}z^{\beta _{2}}\left( 1-z\right) ^{\beta _{3}-\Xi }{F\Bigg(\Xi ,\Xi
-}\widetilde{{\gamma }}{;1+}\widetilde{{\beta }}{;\,{\frac{z}{z-1}}}\Bigg)%
+C_{2}z^{-\beta _{2}}\left( 1-z\right) ^{\beta _{3}-\widehat{\Xi }}{F\Bigg(%
\widehat{\Xi },\widehat{\Xi }-}\widetilde{{\gamma }}{;1{-}\widetilde{{\beta }%
};\,{\frac{z}{z-1}}}\Bigg),  \label{66nnn}
\end{equation}

where

\begin{equation}
{\widehat{\Xi }=}\Xi \left( \widetilde{{\beta }}\rightarrow -\widetilde{{%
\beta }}\right) .  \label{67n}
\end{equation}

If one changes the independent variable $z$ to a new variable $u$ via the
following transformation

\begin{equation}
u={{\frac{z}{z-1}=}}\frac{r-r_{+}}{r-r_{-}},\text{ \ \ \ }\rightarrow \text{%
\ \ \ \ }z=\frac{u}{u-1}.  \label{68n}
\end{equation}

equation (66) recasts in

\begin{equation}
R(u)=C_{1}u^{\widehat{\alpha }}(1-u)^{\widehat{\beta }}F(\widehat{a},%
\widehat{b},\widehat{c};u)+C_{2}u^{-\widehat{\alpha }}(1-u)^{\widehat{\beta }%
}F(\widehat{a}-\widehat{c}+1,\widehat{b}-\widehat{c}+1,2-\widehat{c};u).
\label{69n}
\end{equation}

where 
\begin{eqnarray}
\widehat{\alpha } &=&-i\frac{\omega r_{0}r_{+}-ma}{r_{+}-r_{-}},  \notag \\
\widehat{\beta } &=&\frac{1}{2}-\sqrt{\lambda ^{2}+\frac{1}{4}-\omega
^{2}r_{0}^{2}},  \label{70nn}
\end{eqnarray}

and

\begin{eqnarray}
\widehat{a} &=&\widehat{\beta }-i\omega r_{0},  \notag \\
\widehat{b} &=&\widehat{\beta }-i\frac{\omega r_{0}(r_{+}+r_{-})-2ma}{%
r_{+}-r_{-}},  \notag \\
\widehat{c} &=&1-\frac{2i(\omega r_{0}r_{+}-ma)}{r_{+}-r_{-}}.  \label{71nnn}
\end{eqnarray}

Meanwhile, in the absence of charge and mass the angular equation (29)
admits a normalizable solution when it is expressed in term of the
spheroidal harmonics \cite{MyBook} with eigenvalues $\lambda =-l(l+1)$ in
which $l$ denotes the orbital quantum number. In fact, radial solution (69)
was thoroughly studied by Li \cite{kg12} several years ago. From now on, we
review the computations of \cite{kg12} to obtain the greybody factor of the
RLDBH.

For studying the absorption features of black holes, one should consider two
boundary conditions: pure ingoing modes near the horizon and both ingoing
and outgoing modes at spatial infinity. Following \cite{kg12}, one can infer
from the ingoing boundary condition of the horizon (no outgoing wave
survives at the horizon), the coefficient $C_{2}$ in Eq. (69) must be
vanished. Thus, considering one of the hypergeometric transformations [see
equation (15.3.6) of \cite{MyBook}] we can obtain the asymptotic behavior of
the radial solution as 
\begin{equation}
R(r)=C_{1}\left[ \left( \frac{r}{r_{+}-r_{-}}\right) ^{-\widehat{\beta }}%
\frac{\Gamma (\widehat{c}-\widehat{a}-\widehat{b})\Gamma (\widehat{c})}{%
\Gamma (\widehat{c}-\widehat{b})\Gamma (\widehat{c}-\widehat{a})}+\left( 
\frac{r}{r_{+}-r_{-}}\right) ^{\widehat{\beta }-1}\frac{\Gamma (\widehat{a}+%
\widehat{b}-\widehat{c})\Gamma (\widehat{c})}{\Gamma (\widehat{a})\Gamma (%
\widehat{b})}\right] .  \label{72nn}
\end{equation}

On the other hand, the radial equation (27) with $\mu _{s}=q=0$ near spatial
infinity of the RLDBH is reduced to the following simple second order
differential equation: 
\begin{equation}
r^{2}\frac{d^{2}R}{dr^{2}}+2r\frac{dR}{dr}+\left[ \omega ^{2}r_{0}^{2}-l(l+1)%
\right] R=0,  \label{73nn}
\end{equation}

The solution of the above equation is given by 
\begin{equation}
R(r)=D_{1}r^{-\widehat{\beta }}+D_{2}r^{\widehat{\beta }-1}.  \label{74nn}
\end{equation}%
When $\omega r_{0}>l+1/2$ (high-energy mode), Eq. (74) can also be expressed
as a complex solution

\begin{equation}
R(r)=\frac{1}{\sqrt{r}}\left( D_{1}e^{i\sigma \ln {r}}+D_{2}e^{-i\sigma \ln {%
r}}\right) ,  \label{75nn}
\end{equation}

where

\begin{equation}
\sigma =\sqrt{\omega ^{2}r_{0}^{2}-\left( l+1/2\right) ^{2}}.  \label{76nn}
\end{equation}

It is obvious from Eq. (75) that the first term represents the outgoing wave
while the second term stands for the ingoing wave. Comparing the asymptotic
solutions (72) and (74), one can get the following relationships between the
coefficients 
\begin{eqnarray}
D_{1} &=&C_{1}(r_{+}-r_{-})^{\widehat{\beta }}\frac{\Gamma (\widehat{c}-%
\widehat{a}-\widehat{b})\Gamma (\widehat{c})}{\Gamma (\widehat{c}-\widehat{b}%
)\Gamma (\widehat{c}-\widehat{a})},  \notag \\
D_{2} &=&C_{1}(r_{+}-r_{-})^{1-\widehat{\beta }}\frac{\Gamma (\widehat{a}+%
\widehat{b}-\widehat{c})\Gamma (\widehat{c})}{\Gamma (\widehat{a})\Gamma (%
\widehat{b})}.  \label{77n}
\end{eqnarray}

The conserved flux is given by 
\begin{equation}
\mathcal{F}=-\frac{i}{2}\sqrt{-g}g^{rr}\left( R^{\ast }\partial
_{r}R-R\partial _{r}R^{\ast }\right) .  \label{78n}
\end{equation}%
After substituting the asymptotic solution (75) of the high-energy modes
into Eq. (78), we get the asymptotic flux as follows 
\begin{equation}
\mathcal{F}^{asy}=\frac{\sigma }{2}\left( |D_{1}|^{2}-|D_{2}|^{2}\right) ,
\label{79n}
\end{equation}

Thus, one can remark that $\frac{1}{2}\sigma |D_{1}|^{2}$ and $\frac{1}{2}%
\sigma |D_{2}|^{2}$ terms denote the outgoing \ and ingoing fluxes,
respectively. Hence, we can compute the reflection coefficient for the
high-energy modes as 
\begin{eqnarray}
\mathcal{%
\mathbb{R}
} &=&\frac{|D_{1}|^{2}}{|D_{2}|^{2}}=\left\vert \frac{\Gamma \left[ \frac{1}{%
2}-i\left( \sigma +\Bbbk \right) \right] \Gamma \left[ \frac{1}{2}-i\left(
\Bbbk +\omega r_{0}\right) \right] }{\Gamma \left[ \frac{1}{2}+i\left(
\sigma -\Bbbk \right) \right] \Gamma \left[ \frac{1}{2}+i\left( \Bbbk
-\omega r_{0}\right) \right] }\right\vert ^{2},  \notag \\
&=&\frac{\cosh \pi \left( \sigma -\Bbbk \right) \cosh \pi \left( \sigma
-\omega r_{0}\right) }{\cosh \pi \left( \sigma +\Bbbk \right) \cosh \pi
\left( \sigma +\omega r_{0}\right) },  \label{80n}
\end{eqnarray}%
where

\begin{equation}
\Bbbk =\frac{(r_{+}+r_{-})\omega r_{0}-2ma}{2\kappa r_{+}r_{0}}.  \label{81n}
\end{equation}

It is worth noting that $\cosh $ forms of the Gamma functions seen in Eq.
(80) come from the Euler's reflection formulae of the Gamma function \cite%
{MyBook}. The greybody factor or the absorption probability is given by 
\begin{equation}
\gamma _{GB}=1-\mathcal{%
\mathbb{R}
}.  \label{82n}
\end{equation}

One can check that when we increase the frequency of the waves from the
starting value $\frac{l+1/2}{r_{0}}$, the greybody factor $\gamma _{GB}$
rapidly goes to 1: $\lim_{\omega \left( >\frac{l+1/2}{r_{0}}\right)
\rightarrow \infty }\gamma _{GB}\rightarrow 1$ (see also the figure 1
depicted in \cite{kg12}). The particle flux \cite{Nietzke}, in general,
obeys $\mathcal{F}_{p}\left( \omega \right) \propto \frac{\gamma _{GB}}{%
\mbox{e}^{\omega /T}-1}$, where $T$ denotes the temperature. The
interpretation of this expression is that Hawking radiation is produced with
a thermal spectrum at the horizon and then the spacetime curvature between
the horizon and infinity can scatter some of the radiation back down the
black hole. Thus, the observer at spatial infinity could detect a
non-thermal radiation. On \ the other hand, when $\gamma _{GB}\rightarrow 1$%
, the spectrum of Hawking radiation observed by an asymptotic observer is
pure thermal (black-body radiation). Namely, the highly energetic thermal
waves can overpass the gravitational barrier located at the outside of the
RLDBH and travel from the horizon to the asymptotic region. As a final
remark, it should be kept in mind that the identification of the ingoing and
outgoing fluxes at spatial infinity is governed by the parameter $\widehat{%
\beta }$. According to Eq. (70), $\widehat{\beta }$ is real for the
low-energy modes $\omega <\frac{r_{0}}{2}$ and complex for the high-energy
modes $\omega >\frac{l+1/2}{r_{0}}$. We have shown above that high-energy
modes render possible the calculation of the greybody factor. However, in
the case of the low-energy modes ($\omega <\frac{r_{0}}{2}$) the
identification of the fluxes becomes a very hard task. This is in fact due
to the non-asymptotically flat structure of the RLDBH. This issue was also
discussed in detail in \cite{kg12}.

Now, we want to employ the DRS method to investigate the Hawking radiation
of the RLDBH. Since we now know that the chargeless and massless scalar
(thermal) waves with $\omega >\frac{l+1/2}{r_{0}}$ and $\gamma
_{GB}\rightarrow 1$ can smoothly reach to the observer at spatial infinity,
the radial solution (51) with $q=\mu _{s}=0$ can be used for the application
of the DRS method. From Eq. (52), we can see that Eq. (51) near the exterior
event horizon ($r\rightarrow r_{+}$, $z\rightarrow 0$) behaves as 
\begin{equation}
R(z)\sim C_{1}z^{^{\beta _{2}^{0}}}+C_{2}z^{^{-\beta _{2}^{0}}}.  \label{83}
\end{equation}

where $\beta _{2}^{0}=\left. \beta _{2}\right\vert _{q=\mu _{s}=0}$. Thus,
the near horizon wave solution can be approximated to 
\begin{equation}
\Psi \sim \mbox{e}^{-i\omega t}z^{^{^{\pm \beta _{2}^{0}}}}.  \label{84}
\end{equation}%
Taking cognizance of Eqs.~(16), (18), and (44), the parameter $\beta
_{2}^{0} $ reads 
\begin{equation}
\beta _{2}^{0}=\frac{i}{2\kappa }\left[ \omega -m\Omega _{H}\right] .
\label{85}
\end{equation}%
Moreover, if we set

\begin{equation}
\varpi =\omega -m\Omega _{H},  \label{86}
\end{equation}

equation (85) can be rewritten as

\begin{equation}
\beta _{2}^{0}=i\frac{\varpi }{2\kappa }.  \label{87}
\end{equation}

Performing Taylor series expansion, we find the structure of the metric
function $f$ around the event horizon to be

\begin{eqnarray}
&&f_{EH}\simeq \left. \frac{df}{dr}\right\vert
_{r=r_{+}}(r-r_{+})+O(r-r_{+})^{2},  \notag \\
&\simeq &2\kappa \left( r_{+}-r_{-}\right) x,  \label{88}
\end{eqnarray}

where $x=-z=\frac{r-r_{+}}{r_{+}-r_{-}}$. Thus, we can express the tortoise
coordinate ($r^{\ast }$) \cite{Chandra} near the horizon as 
\begin{equation}
r^{\ast }\simeq \int \frac{dr}{f_{EH}}=\int \frac{dx}{2\kappa x}=\frac{1}{%
2\kappa }\ln x\approx \frac{1}{2\kappa }\ln \left( r-r_{+}\right) ,
\label{89}
\end{equation}

which corresponds to%
\begin{equation}
r-r_{+}\simeq e^{2\kappa r_{\ast }}.  \label{90}
\end{equation}

Therefore, the ingoing and outgoing wave solutions on the black hole event
horizon surface become

\begin{equation}
\Psi _{in}=\mbox{e}^{-i\omega t}e^{-i\varpi r_{\ast }},  \label{91}
\end{equation}%
\begin{equation}
\Psi _{out}(r>r_{+})=\mbox{e}^{-i\omega t}e^{i\varpi r_{\ast }},  \label{92}
\end{equation}

respectively. To reveal the features of the waves near the event horizon, we
first define 
\begin{equation}
\widetilde{r}=\frac{\varpi }{\omega }r_{\ast },  \label{93}
\end{equation}

and then introduce the Eddington-Finkelstein coordinate:

\begin{equation}
v=t+\widetilde{r}.  \label{94}
\end{equation}

Hence, the ingoing and outgoing wave solutions become 
\begin{equation}
\Psi _{in}=\mbox{e}^{-i\omega \left( t+\widetilde{r}\right) }=\mbox{e}%
^{-i\omega v},  \label{95}
\end{equation}%
\begin{eqnarray}
\Psi _{out}(r>r_{+}) &=&\mbox{e}^{-i\omega \left( t-\widetilde{r}\right) }=%
\mbox{e}^{-i\omega v}\mbox{e}^{2i\varpi r_{\ast }},  \notag \\
&=&\mbox{e}^{-i\omega v}(r-r_{+})^{i\frac{\varpi }{\kappa }}.  \label{96}
\end{eqnarray}

One can easily observe that the $\Psi _{out}(r>r_{+})$ solution is not
analytical at the event horizon and the analytic continuation produces a
damping factor, which makes it possible to attain an expression for the
decay rate $\Gamma _{dy}$ \cite{decay}. To clarify the latter remark, we use
the DRS method by rotating $-\pi $ through the lower-half complex $r$ plane: 
\begin{equation}
(r-r_{+})\rightarrow \left\vert r-r_{+}\right\vert \mbox{e}^{-i\pi
}=(r_{+}-r)\mbox{e}^{-i\pi }.  \label{97}
\end{equation}%
Thus, the outgoing wave solution in the internal region ($r<r_{+}$) yields 
\begin{equation}
\Psi _{out}(r<r_{+})=\mbox{e}^{-i\omega v}(r_{+}-r)^{i\frac{\varpi }{\kappa }%
}\mbox{e}^{\pi \frac{\varpi }{\kappa }}.  \label{98}
\end{equation}

Therefore, Eqs. (96) and (98) represent the analytically continuous outgoing
wave propagating around the RLDBH's event horizon. Thus, the emission rate
or\ the relative scattering probability \cite{kg15} of the scalar wave at
the surface of the event horizon surface becomes 
\begin{equation}
\Gamma _{dy}=\left\vert \frac{\Psi _{out}(r>r_{+})}{\Psi _{out}(r<r_{+})}%
\right\vert ^{2}=\mbox{e}^{-2\pi \frac{\varpi }{\kappa }}.  \label{99}
\end{equation}

Following the other DRS applications like \cite%
{DRS1,DRS11,DRS2,DRS3,DRS4,DRS5,DRS6}, the Hawking radiation spectrum of the
scalar particles emitted from a black hole is obtained via the normalization
condition as follows

\begin{equation}
\left\vert N\left( \varpi \right) \right\vert ^{2}=\frac{\Gamma _{dy}}{%
1-\Gamma _{dy}}=\frac{1}{\mbox{e}^{2\pi \frac{\varpi }{\kappa }}-1}.
\label{100}
\end{equation}

where $N\left( \varpi \right) $\ is the normalization constant. Equation
(100) suggests that the emission of scalar particles has a thermal character
analogous to the well-known blackbody spectrum with temperature $T_{H}=\frac{%
\kappa }{2\pi }$, which is fully agree with Eq. (17).

It is also worth noting that since the rotating dilaton black hole contains
an ergosphere outside the horizon, it might have a superradiant instability.
In asymptotically flat spacetimes, such as Kerr black holes, the
superradiance shows itself as an (quantum) emission of certain modes
ejaculating the angular momentum of the black hole to spatial infinity. On
the other hand, when the considered spacetime has a non-asymptotically flat
geometry, the asymptotic behavior of the modes should be carefully analyzed.
As it was shown in \cite{Clem1,rbh5}, the superradiant modes of the RLDBH do
not propagate to spatial infinity. Instead, they are confined in a region
outside the event horizon of the RLDBH. So that we have an exponential
growth for the superradiant modes. However, this gives rise to the classical
instability of the RLDBH solution. Besides, $\left\vert N\left( \varpi
\right) \right\vert ^{2}$ would diverge. Therefore, in addition to the
condition of $\omega >\frac{l+1/2}{r_{0}}$, the energies/frequencies of the
chargeless and massless scalar fields performing the Hawking emission should
also be higher than the threshold \cite{Aliev} frequency ($m\Omega _{H}$) of
the superradiance: $\omega >m\Omega _{H}.$

\section{Conclusion}

We have presented complete analytical solutions to the covariant KGE for a
charged massive scalar field in the RLDBH spacetime. Both the angular and
radial exact solutions are demonstrated in terms of the confluent Heun
functions \cite{heun0,heun1,heun2,heun3}, and they cover the whole range of
the observable space $0\leq z<\infty $.

In particular, the radial solution has enabled us to analyze the resonant
frequencies of the RLDBH with the help of the $\widetilde{\delta }_{n}$%
-condition \cite{DRS6,heun3,Fiziev1}. We have used the resonant frequencies
to study the spectroscopy of the RLDBH. Both entropy and area spectra are
found to be equally spaced and independent of the RLDBH parameters.
Therefore, our results support the conjecture of Bekenstein \cite{Bek005}.
Then, we have studied the greybody factor problem of the RLDBH in order to
explore which waves can propagate from the horizon to the spatial infinity.
While doing this computation, the lack of the inverse transformation of the
confluent Heun functions which is essential to find the exact asymptotic
form of the radial solution has enforced us to consider the case of
chargeless and massless scalar fields, which was previously considered in 
\cite{kg12,rbh2,rbh3,rbh5}. Using a particular transformation between the
confluent Heun and hypergeometric functions, we have expressed the radial
solution in terms of the hypergeometric functions which posses a broad
spectrum of linear transformation features \cite{MyBook}. We then have
obtained the ingoing and the outgoing fluxes at spatial infinity for the
high-energy modes $\left( \omega >\frac{l+1/2}{r_{0}}\right) $ that admit
the identification of the fluxes, asymptotically. On the other hand, the
low-energy modes ($\omega <\frac{r_{0}}{2}$) did not let us to make the
distinction the ingoing and the outgoing fluxes at spatial infinity; we have
simply ignored that case. Afterwards, we have performed an analytical
computation of the greybody factor. From the limit of $\lim_{\omega \left( >%
\frac{l+1/2}{r_{0}}\right) \rightarrow \infty }\gamma _{GB}\rightarrow 1,$
we have deduced that the high-energetic thermal waves can pass over the
gravitational barriers and move from the horizon to the observer located at
asymptotic region. For those high-energy modes, we have considered the DRS
method, which is a powerful mathematical tool of the analytic continuation
for identifying the Hawking temperature of the considered black hole. To
this end, we have used the series expansion of the confluent Heun function
[see Eq. (52)] and obtained the outgoing wave solution in the exterior and
interior regions of the event horizon. Finally, Hawking radiation spectrum
of the scalar particles emitted from the RLDBH was derived, and Hawking
temperature (17) of the RLDBH was successfully obtained.

The results of the study are promising and motivate further work in this
direction. In particular, the results can be extended to other particles
with nonzero spin and to the black holes of higher dimensions.

\section*{Acknowledgments}

The author is grateful to the editor and anonymous referees for their
valuable comments and suggestions to improve the paper.

\bigskip

\newpage

{\LARGE Appendix}

\bigskip

For reference, the following is a list of symbols that are used often
throughout the text.

\begin{tabular}{|l|l|}
\hline
$M$, $M_{QL}$ & Mass parameter of the RLDBH spacetime and quasilocal mass ($%
M_{QL}=M/2$)$.$ \\ \hline
$\mathcal{A}$; $\phi ,$ $\chi ,$ & Electromagnetic vector potential; Dilaton
and axion fields. \\ \hline
$Q$, $r_{0}$ & Background electric charge and charge parameter ($r_{0}=\sqrt{%
2}Q$) . \\ \hline
$J$, $a$, $\Omega _{H}$ & Angular momentum, rotation parameter: $a=2J/r_{0}$ 
$\in $ $[0,M]$, and angular velocity ($\Omega _{H}=\frac{a}{r_{+}r_{0}}$).
\\ \hline
$r_{+}$, $r_{-}$ & Outer horizon and inner horizon. \\ \hline
$\kappa $ , $T_{H}$ & Surface gravity and Hawking temperature ($T_{H}=\frac{%
\hslash \kappa }{2\pi }$). \\ \hline
$A_{BH}$ , $S_{BH}$ & Black hole area and entropy. \\ \hline
$\Psi ;$ $\mu _{s}$ , $q$ & Scalar field (wavefunction); Mass and charge
parameter of the scalar field. \\ \hline
$\omega $ & Frequency. The time dependence of any field is $\sim $ $%
e^{-i\omega t}$. \\ \hline
$n$ & Overtone numbers of the eigenfrequencies. It starts from a fundamental
mode with $n=0$. \\ \hline
$\omega _{n},$ $\Delta \omega $ & Resonant (quasinormal mode) frequency and
transition frequency: $\Delta \omega \approx $\textit{Im}$\omega _{n-1}-$%
\textit{Im}$\omega _{n}.$ \\ \hline
$m,$ $\lambda $ & Azimuthal number with respect to the axis of rotation and
eigenvalue. \\ \hline
$\Phi _{e},$ $I_{adb}$ & Electric potential of the RLDBH and adiabatic
invariant quantity. \\ \hline
$\widetilde{\alpha },\widetilde{\beta },\widetilde{\gamma },\widetilde{%
\delta },\widetilde{\eta }$ & Parameters of the confluent Heun's function ($%
HeunC$). \\ \hline
$y,$ $z$ & Independent variables of the confluent Heun's function: $y=\frac{%
1-\cos \left( \theta \right) }{2}$ and $z=\frac{r-r_{+}}{r_{-}-r_{+}}.$ \\ 
\hline
$\widetilde{\mu }$, $\widetilde{\nu }$ & $\widetilde{\mu }=\frac{1}{2}(%
\widetilde{\alpha }-\widetilde{\beta }-\widetilde{\gamma }+\widetilde{\alpha 
}\widetilde{\beta }-\widetilde{\beta }\widetilde{\gamma })-\widetilde{\eta }$
and $\widetilde{\nu }=\frac{1}{2}(\widetilde{\alpha }+\widetilde{\beta }+%
\widetilde{\gamma }+\widetilde{\alpha }\widetilde{\gamma }+\widetilde{\beta }%
\widetilde{\gamma })+\widetilde{\delta }+\widetilde{\eta }.$ \\ \hline
$\widehat{a},$ $\widehat{b},$ $\widehat{c}$ & Parameters of the
hypergeometric function. \\ \hline
$u$ & Independent variables of the hypergeometric function: $u={{\frac{z}{z-1%
}=}}\frac{r-r_{+}}{r-r_{-}}.$ \\ \hline
$\mathcal{%
\mathbb{R}
}$ , $\gamma _{GB}$ & Reflection coefficient and greybody factor ($\gamma
_{GB}=1-\mathcal{%
\mathbb{R}
}$). \\ \hline
$r_{\ast }$, $\varpi $ & Tortoise coordinate and wave frequency detected by
the observer rotating with \\ 
& the horizon: $\varpi =\omega -m\Omega _{H}.$ \\ \hline
$\Gamma _{dy},$ $N\left( \varpi \right) $ & Emission rate and normalization
constant. \\ \hline
$l$ & Integer angular number, related to the eigenvalue $\lambda =-l(l+1)$.
\\ \hline
$\Psi _{in},$ $\Psi _{out}$ & Ingoing and outgoing wave solutions around the
event horizon. \\ \hline
$\mathcal{F}$, $\mathcal{F}^{asy}$, $\mathcal{F}_{p}$ & Conserved flux,
asymptotic flux, and particle flux. \\ \hline
$C_{1}$, $C_{2}$, $D_{1}$, $D_{2}$ & Integral constants of the wave
solutions. \\ \hline
$q^{\ast },$ $\tau ,$ $\beta _{2}^{0},$ $\widetilde{r}$ & $q^{\ast }=\frac{q%
}{\sqrt{2}}$, $\tau =2aq^{\ast }$, \ $\beta _{2}^{0}=i\frac{\varpi }{2\kappa 
},$ and $\widetilde{r}=\frac{\varpi }{\omega }r_{\ast }.$ \\ \hline
$\Xi $, ${\widehat{\Xi }}$ & $\Xi =\frac{1+\widetilde{\beta }+\widetilde{{%
\gamma }}{+}\sqrt{\widetilde{{\beta }}^{2}+\widetilde{{\gamma }}^{2}+1-4%
\widetilde{\eta }}}{2}$ and ${\widehat{\Xi }=}\Xi \left( \widetilde{{\beta }}%
\rightarrow -\widetilde{{\beta }}\right) .$ \\ \hline
$\sigma ,$ $\Bbbk $ & $\sigma =\sqrt{\omega ^{2}r_{0}^{2}-\left(
l+1/2\right) ^{2}}$ and $\Bbbk =\frac{(r_{+}+r_{-})\omega r_{0}-2ma}{2\kappa
r_{+}r_{0}}.$ \\ \hline
$\widetilde{\delta }_{n}$-condition & $1+\frac{\widetilde{\beta }+\widetilde{%
\gamma }}{2}+\frac{\widetilde{\delta }}{\widetilde{\alpha }}=-n.$ \\ \hline
\end{tabular}

\end{document}